\documentclass[aps,prb,twocolumn,twoside,floatfix,superscriptaddress]{revtex4}
\usepackage{graphicx}
\usepackage{graphics}
\usepackage{amsmath}
\usepackage{amssymb}
\usepackage{mathrsfs}
\usepackage{epsf}

\begin{document}

\title{Microscopic realization of cross-correlated noise processes}

\author{Anindita Shit}
\affiliation{Department of Chemistry, Bengal Engineering and Science University,
Shibpur, Howrah 711103, India}

\author{Sudip Chattopadhyay }
\altaffiliation{Corresponding author}
\email{sudip_chattopadhyay@rediffmail.com}
\affiliation{Department of Chemistry, Bengal Engineering and Science University,
Shibpur, Howrah 711103, India}

\author{Suman Kumar Banik}
\email{skbanik@bic.boseinst.ernet.in}\affiliation{Department of
Chemistry, Bose Institute, 93/1 A P C Road, Kolkata 700009, India}

\author{Jyotipratim Ray Chaudhuri}
\altaffiliation{Corresponding author}
\email{jprc_8@yahoo.com}
\affiliation{Department of Physics, Katwa College, Katwa, Burdwan 713130, India}

\date{\today}

\begin{abstract}
We present a microscopic theory of cross-correlated noise
processes, starting from a Hamiltonian system-reservoir
description. In the proposed model, the system is nonlinearly
coupled to a reservoir composed of harmonic oscillators, which in
turn is driven by an external fluctuating force. We show that the
resultant Langevin equation derived from the composite system
(system+reservoir+external modulation) contains the essential
features of cross-correlated noise processes.
\end{abstract}

\maketitle

{\bf In this paper, we deal with a bath that is being nonlinearly
driven by an external noise. This work is an attempt to analyze
the mechanism of the action of additive, as well as multiplicative
cross correlated noises, as well as to get an insight of the
mutual interplay of these noises from a microscopic standpoint.
This is in contrast with the traditional works which essentially
treats the external cross correlated noises in a phenomenological
manner. We point out the important fact that the non-linear
driving becomes ubiquitous to explain (i) the actual microscopic
origin of space-dependent dissipation and multiplicative noise,
and (ii) the origin of external multiplicative cross correlated
noise.}

\section{Introduction}
A huge impetus has been envisaged during the last few decades
towards research leading to an in-depth understanding of the
detailed dynamics of systems that are subjected to external noise
fields. Experimental and theoretical studies in the recent past
have revealed the ubiquitous and constructive role of noises in a
plethora of physical phenomena ranging from the self-organization
and dissipative dynamics of systems, noise-induced transitions,
phase transitions driven by noises, thermal ratchets (or Brownian
motors) to even the stochastic resonance in zero-dimensional and
spatially extended systems. Noises have their own role to play in
a host of problems pertaining to physical, chemical, biological
relevance as well as those related to economics phenomena [see
Refs.\cite{refsc1,refsc2,refsc3,refsc4}]. The common feature of an
overwhelming majority of these studies is that the system is
thermodynamically closed and the energy conservation is guided by
the celebrated fluctuation-dissipation relation\cite{kubobookII}.
However, in a number of situations the system may be
thermodynamically open\cite{pccp5}. There exists no
fluctuation-dissipation relation for the thermodynamically open
system\cite{refsc5} and recently such a system has attained a wide
attention\cite{jcp19,jcp21,jcp32,jcp33,jcp34}. The origin of the
noise in the thermodynamically open system driven by two or more
random forces may be different. The barrier crossing dynamics with
multiplicative and additive noises initiated strong interest in
the early 1980s. Formerly, in most of the initial works, noise
forces that were present simultaneously in stochastic physical
processes have usually been treated as uncorrelated random
variables, since it has assumed that they have different noise
origins. However, there are some situations where noises in some
open systems may have a common origin and hence can be
correlated\cite{pccp6}. If this happens, then the statistical
properties of the noises should not be much different and can be
correlated with each other. It has also been envisaged in many
situations that strong external noises often modify the internal
properties of a dynamical system, thus, in a way, necessitating a
that the internal and external noises be independent of each
other. The case of simultaneous acting uncorrelated and correlated
additive and multiplicative noise has been studied by various
authors. The study of nonlinear dynamical systems perturbed
cross-correlated noises has become an attractive subject in recent
years\cite{refsc6}. Correlated noise processes have made their
presence felt in a wide range of studies, such as the statistical
properties of a single mode laser \cite{pccp9}, bistable
kinetics\cite{pccp10}, barrier crossing dynamics\cite{jcp,pre},
steady state entropy production\cite{pccp13} and noise induced
transport\cite{pccp}. Recent time, it is now well accepted that
the effect of correlation between additive and multiplicative
noise is indispensable in explaining phenomena such as phase
transition, transport of motor protein and the presence of
cross-correlated noises changes the dynamics of the
system\cite{pccp30}.

It was first pointed out by Fedchenia \cite{pccp2}that the
cross-correlated noises have a vital role to play in the realm of
hydrodynamics of vortex flow from a common origin that appear in
the time evolution equation of dimensionless modes of flow rates.
In literature, numerous evidences fortify the effect of the
interference among the additive and multiplicative noises in the
context of dynamics, phase transition and many other relevant
phenomena. While Fulinski and Telejko \cite{pccp3} have studied
the effect of correlation between the additive and multiplicative
white noise in the kinetics of the bistable systems, Madureira et
al's work \cite{pccp27} throws light on the role of the coupled
effect of two correlated white noises (additive and
multiplicative) on the escape-rate of double-well systems. In this
context, the role of correlation between additive and
multiplicative noises have been explored by Mei {\it et al.}
\cite{pccp28}during their study of the relaxation of a bistable
system driven by cross-correlated noises. In one of the important
developments in this direction, H\"{a}nggi and co-workers
\cite{pccp29} have put forth the influence of  two white Gaussian
noise sources that are correlated, with one being associated with
an additive white noise and the other with a multiplicative white
noise.

It has been experienced in dealing with the dynamical systems that
in several cases the effect of adding up noises makes the system
more ordered. For example, this is envisaged in the phenomenal
suppression of activation present in bistable/metastable systems.
It has been observed that in such systems that the lifetime is
markedly prolonged upon the addition of two correlated Gaussian
noise terms, {\it vis- a -vis} the lifetime in the presence of a
single noise term \cite{pccp27}. At the level of a Langevin type
description of a dynamical system\cite{pccp6}, the presence of
correlation between noises can change the dynamics of the
system\cite{pccp9,pccp30,pccp7}. In recent years, there has been
an increasing interest in studying the effects of the noises in
the (nonlinear) dynamical systems.

The aim of the present work is to investigate the microscopic
origin of mutual correlation of additive and multiplicative
external noises. We hope our model can be used as an efficient
tool to study mechanism of various above mentioned effects and
phenomena in systems driven by cross-correlated noises, one
additive and the other multiplicative.

\section{Theoretical Development}


To start with, we consider the system to be coupled to a harmonic
heat bath with characteristic frequency sets $\{ \omega_j\}$. The
coupling between the system and the bath is, in general,
considered to be nonlinear as well. Initially at $t=0$, the bath
is in thermal equilibrium at a temperature $T$. At $t=0_+$, the
external fluctuating force $\epsilon(t)$ is switched on, which
modulates the harmonic heat bath. The Hamiltonian for the
composite system thus can be written as
\begin{equation}
\label{eq1}
H = H_S + H_B + H_{SB} + H_{int},
\end{equation}

\noindent where, $H_S=(p^2/2) + V(q)$ is the system's Hamiltonian
with $q$ and $p$ being the  co-ordinate and momentum of the system
of interest having unit mass. Here, $V(q)$ is the external force
field. The second and third terms in the right hand side of
Eq.(\ref{eq1}) refer to the harmonic heat bath and the interaction
between the system and the latter,
\begin{eqnarray*}
H_B + H_{SB} = \sum_{j=1}^N \left \{
\frac{p_j^2}{2} + \frac{1}{2} \omega_j^2 \left(q_j- c_jf(q) \right)^2
\right \},
\end{eqnarray*}

\noindent with $q_j$ and $p_j$ being the bath variables. The
quantity $c_j$ is the coupling constant and $f(q)$ is a smooth
well behaved function of system variables. The last term in
Eq.(\ref{eq1}), $H_{int}$, takes care of the interaction between
the harmonic bath and the external fluctuations $\epsilon(t)$
\begin{equation}\label{eq2}
H_{int} = \sum_{j=1}^N \kappa_j h(q_j) \epsilon(t),
\end{equation}

\noindent with, $\kappa_j$ being the strength of the interaction
and $h(q_j)$ is an arbitrary analytical function of bath variables
and in general, nonlinear.
The external driving force $\epsilon(t)$ is considered
to be a stationary delta correlated one with unit noise strength and follows the
Gaussian statistics
\begin{eqnarray}\label{eq3}
\langle \epsilon(t)\rangle = 0,   \langle \epsilon(t)
\epsilon(t^{\prime})\rangle = 2 \delta(t-t^{\prime}).
\end{eqnarray}

Now, from Eq.(\ref{eq1}) the various equations of motion for the
system and bath variables are found to be
\begin{eqnarray}\label{eq4}
\dot{q} &=& p , \;
\dot{p} = - V^{\prime}(q) +
\sum_{j=1}^N c_j \omega_j^2 \left(q_j- c_jf(q) \right) f^{\prime}(q), \\
\label{eq5}
\dot{q_j} &=& p_j, \;
\dot{p_j} = - \omega_j^2
\left(q_j- c_j f(q) \right) - \kappa_j \frac{dh}{dq_j} \epsilon(t).
\end{eqnarray}

\noindent Using the explicit form of the function $h(q_j)$
\begin{eqnarray}\label{eq5a}
 h(q_j)=q_j+\frac{1}{2}q_j^2,
 \end{eqnarray}

\noindent Eq.(\ref{eq5}) becomes
\begin{eqnarray}\label{eq6}
\ddot{q_j} + \left \{ \omega_j^2 + \kappa_j \epsilon(t)\right \}
q_j =c_j \omega_j^2 f(q) + \kappa_j \epsilon(t) .
\end{eqnarray}

\noindent An analysis of Eq.(\ref{eq6}) reveals the fact that the
presence of the term $\epsilon(t)$ in the equation essentially
transforms an otherwise simple harmonic oscillator driven by a
force [the terms on the right had side of Eq.(\ref{eq6})] to a
situation with an oscillator having fluctuating (or random)
frequencies. While in the absence of $\epsilon(t)$, we could have,
at least in principle, sought for the analytic solution of the
problem. In the present situation, needless to mention is the fact
that we have to resort to some approximation method to solve
Eq.(\ref{eq6}). The standard system- reservoir model assumes that
the any change of the system degrees of freedom does not affect
the spatio-temporal evolution of the harmonic bath, while the
reverse is not true \cite{lindenberg}. As a consequence, we use
the perturbative solution of the bath to monitor the change in the
system and thereby eliminate the bath variables from the system
description. In nut-shell, we resort to a standard approximation
for the solution of Eq.(\ref{eq6}) in which the harmonic baths
remain unaffected by the system, so that we neither need to seek
for a simultaneous solution of the system and bath variables, nor
any explicit perturbative correction is needed for them.

\noindent To solve Eq.(\ref{eq6}) we assume
\begin{eqnarray}\label{eq7}
q_j(t) = q_j^0(t) + \kappa_j q_j^{1}(t) ,
\end{eqnarray}

\noindent with $|\kappa_j|<1$ [$\kappa_j$ is small, and of the
same order for all $j$ ] and both $q_j^{0}(t)$  and $q_j^{1}(t)$
[representing a small perturbation around $q_j^{0}(t)$] satisfy
the equations
\begin{eqnarray}\label{eq8}
\ddot{q}_j^0(t) + \omega_j^2 q_j^0(t) = c_j \omega_j^2 f[q(t)] , \\
\label{eq9} {\rm and } \;\;  \ddot{q}^{1}(t) + \omega_j^2
q_j^{1}(t) = - q_j^0(t) \epsilon(t) - \epsilon(t) ,
\end{eqnarray}

\noindent respectively. Eqs.(\ref{eq8},\ref{eq9}) have been
written on the physical ground that at $t=0$ the heat bath is in
thermal equilibrium, which in turn is modulated by external
fluctuations $\epsilon(t)$, at $t=0_+$. Under such condition,
$q_j(0) = q_j^0(0)$ and $p_j(0) = p_j^0(0)$. This also implies
that $q_j^{1}(0) = p_j^{1}(0) =0 $. Making use of the formal
solution(s) of Eq.(\ref{eq8}), namely,
\begin{eqnarray}\label{eq10}
q_j^0(t) &=& q_j^0(0) \cos \omega_j t + \frac{p_j^0(0)}{\omega_j}\sin \omega_j t \nonumber \\
&& + c_j \omega_j \int_0^t dt^{\prime} \sin \omega_j
(t-t^{\prime}) f[q(t^{\prime})],
\end{eqnarray}

\noindent we obtain the following expression for the bath variable
$q_j(t)$
\begin{eqnarray}\label{eq11}
q_j(t) &=& c_j f(q) + \left \{ q_j^0(0) - c_j f[q(0)] \right \}
\cos \omega_j t
+ \frac{p_j^0(0)}{\omega_j}\sin \omega_j t \nonumber \\
&& - c_j \int_0^t dt^{\prime} \cos \omega_j (t-t^{\prime}) f^{\prime}
[q(t^{\prime})] \dot q(t^{\prime})  \nonumber \\
&& - \frac{\kappa_j}{\omega_j} \int_0^t dt^{\prime}\sin\omega_j
(t-t^{\prime}) \epsilon(t^{\prime}) \left [ 1 + q_j^0 (t^{\prime})
\right ] .
\end{eqnarray}

Now inserting Eq.(\ref{eq11}) into Eq.(\ref{eq4}), we obtain the
dynamical equation for the system variable as
\begin{eqnarray}\label{eq12}
\dot q &=& p, \nonumber \\
\dot p &=& - V^{\prime}(q) -
f^{\prime}(q) \int_0^t dt^{\prime} \gamma(t-t^{\prime})
f^{\prime}[q(t^{\prime})] p(t^{\prime}) \nonumber \\
&&- f^{\prime}(q) \sum_{j=1}^N c_j \kappa_j \omega_j
\int_0^t dt^{\prime} \sin\omega_j (t-t^{\prime})
\epsilon(t^{\prime}) q_j^0(t^{\prime}) \nonumber \\
&& + f^{\prime}(q(t)) \xi(t) + f^{\prime}[q(t)] \pi(t)
\end{eqnarray}

\noindent where the random force $\xi(t)$ and the memory kernel
$\gamma(t)$ are given by
\begin{eqnarray}\label{eq13}
\xi(t) &=& \sum_{j=1}^N c_j \omega_j^2 \left[ \{ q_j^0(0) -c_j
f[q(0)] \} \cos \omega_j t  \right. \nonumber \\
&& \left. + \frac{p_j(0)} {\omega_j} \sin \omega_j t\right], \\
\label{eq14}
\gamma(t) &=& \sum_{j=1}^N c_j^2 \omega_j^2 \cos \omega_j t .
\end{eqnarray}

\noindent In Eq.(\ref{eq12}), $\pi(t)$ is the fluctuating force generated
due to the linear part of the coupling function $h(q_j)$ and is
given by
\begin{eqnarray}\label{eq15}
\pi(t) = - \int_0^t dt^{\prime} \varphi(t-t^{\prime})
\epsilon(t^{\prime}) ,
\end{eqnarray}
\noindent where
\begin{eqnarray}\label{eq16}
\varphi(t) = \sum_{j=1}^N c_j \omega_j \kappa_j \sin(\omega_j t) .
\end{eqnarray}

The form of Eq.(\ref{eq12}) reflects that the system dynamics is
effectively modulated by three fluctuating forces, $\xi(t)$, the
internal thermal noise and the other two fluctuating forces due to
driving of the bath by an external random force $\epsilon(t)$. If
the external noise-bath coupling is linear, i.e., when
$h(q_j)=q_j$, the fourth term disappears and we obtain our earlier
results \cite{jcp21,endb1,endb2}. It is important to note that all
the effective fluctuating forces are multiplicative in nature due
to the presence of the system variable function $f^{\prime}(q)$.
The statistical properties of $\xi(t)$ can be derived by using
suitable canonical thermal distribution of bath coordinates and
momenta at $t=0$. We assume that the bath variables $\{ q_j(0),
p_j(0) \}$ are distributed according to the Gaussian form with the
probability distribution function, $W\{ q_j(0), p_j(0)\} = (1/Z)
\exp (- [H_B+H_{SB}]/k_BT)$ where $Z$ is the partition function.
Then, the statistical properties of the fluctuating force $\xi(t)$
become $\langle \xi(t) \rangle = 0 , \langle \xi(t)
\xi(t^{\prime}) \rangle = 2 \gamma (t-t^{\prime})k_BT$.

To identify Eq.(\ref{eq12}) as a {\it generalized Langevin
equation } for nonequilibrium open system, we need to impose some
conditions on the coupling co-efficients $c_j$ and $\kappa_j$, on
the bath frequencies $\omega_j$ and on the number $N$ of the bath
oscillators that ensure $\gamma(t)$ to be indeed dissipative. A
sufficient condition for $\gamma(t)$ to be dissipative is that it
is positive definite and decreases monotonically with time. These
conditions are achieved if $N\rightarrow \infty$ and if $c_j
\omega_j^2$ and $\omega_j$ are sufficiently smooth functions of
$j$ \cite{fordkacmajur}. As $N\rightarrow \infty$ one replaces the
sum by an integral over $\omega$ weighted by a density of state
$\rho(\omega)$. Thus, to obtain a finite result in the continuum
limit, the coupling function $c_j=c(\omega)$ and
$\kappa_j=\kappa(\omega)$ are chosen as
\begin{eqnarray}\label{eq17a}
c(\omega) = \frac{c_0}{\omega{\sqrt \tau_c}}, \;\; {\rm and }\;\;
\kappa(\omega) = \kappa_0 \omega {\sqrt \tau_c} .
\end{eqnarray}

\noindent Consequently, $\gamma(t)$ and $\varphi(t)$ reduce to
\begin{eqnarray}\label{eq18}
\gamma(t) &=&  \frac{c_0^2}{\tau_c} \int_0^{\infty} d\omega \rho
(\omega) \cos(\omega t) \\
\label{eq19} {\rm and }\;\; \varphi(t) &=&  c_0 \kappa_0
\int_0^{\infty} d\omega \rho (\omega) \omega \sin(\omega t) ,
\end{eqnarray}

\noindent where $c_0$ and $\kappa_0$ are constants and
$\omega_c=1/\tau_c$ is the cut-off frequency of the bath
oscillators. $\tau_c$ may be regarded as the correlation time of
the bath oscillators and $\rho(\omega)$ is the density of modes of the
heat bath which is assumed to be Lorentzian:
$\rho(\omega) = 2 \tau_c/[\pi (1+ \omega^2 \tau_c^2)]$ .
With these forms of $\rho(\omega)$, $c(\omega)$ and
$\kappa(\omega)$, $\gamma(t)$ and $\varphi(t)$ take the following
forms:
\begin{eqnarray}\label{eq21}
\gamma(t) &=& \frac{c_0^2}{\tau_c} \exp (-|t| / \tau_c)
                      = \frac{\gamma}{\tau_c} \exp (-|t| / \tau_c) , \\
\label{eq22}
\varphi(t) &=& \frac{c_0 \kappa_0}{\tau_c} \exp (-|t| / \tau_c) .
\end{eqnarray}

\noindent For $\tau_c\rightarrow0$, Eqs.(\ref{eq21},\ref{eq22})
become $\gamma(t)=2\gamma\delta(t)$, where $\gamma=c_0^2$ and
$\varphi(t)=2c_0\kappa_0\delta(t)$; and consequently, one obtains
the $\delta$-correlated noise processes describing Markovian
dynamics.

The third term on the right hand side of Eq.(\ref{eq12}) can be
written as
\begin{eqnarray} \label{neweq1}
&& f^\prime\left[q(t)\right]\int_0^t dt^\prime \epsilon(t^\prime)
\int_0^{t^\prime} dt^{\prime\prime} \left[\left\{\sum_j c_j^2
\kappa_j \omega_j^2 \sin\omega_j (t-t^\prime) \right. \right. \nonumber \\
&& \left. \left. \times \sin\omega_j
(t^\prime-t^{\prime\prime})\right\}f\left[q(t^{\prime\prime})\right]
\right]
\end{eqnarray}

\noindent where we have used the particular solution for $q_j^0(t)$ [see
Eq.(\ref{eq10})]. The above expression can be rewritten, using
known trigonometric identity, as
\begin{eqnarray}\label{neweq2}
&&\frac{1}{2}f^\prime\left[q(t)\right]\int_0^t dt^\prime
\epsilon(t^\prime) \int_0^{t^\prime} dt^{\prime\prime}
f\left[q(t^{\prime\prime})\right]\sum_j c_j^2 \kappa_j \omega_j^2 \nonumber \\
&& \times \cos\omega_j (t-2t^\prime+t^{\prime\prime})\nonumber \\
&&-\frac{1}{2}f^\prime\left[q(t)\right]\int_0^t dt^\prime
\epsilon(t^\prime) \int_0^{t^\prime} dt^{\prime\prime}
f\left[q(t^{\prime\prime})\right]\sum_j c_j^2 \kappa_j \omega_j^2 \nonumber \\
&& \times \cos\omega_j (t-t^{\prime\prime})
\end{eqnarray}

\noindent Now, in the continuum limit, and for Markovian internal
dissipation, one easily obtains that
\begin{eqnarray}\label{neweq3}
\sum_j c_j^2 \kappa_j \omega_j^2 \cos\omega_j
(t-2t^\prime+t^{\prime\prime})=2c_0^2 \kappa_0\delta
(t-2t^\prime+t^{\prime\prime})
\end{eqnarray}
and
\begin{eqnarray}\label{neweq4}
\sum_j c_j^2 \kappa_j \omega_j^2 \cos\omega_j
(t-t^{\prime\prime})=2c_0^2 \kappa_0\delta (t-t^{\prime\prime})
\end{eqnarray}
Now keeping in mind that $t > t^\prime > t^{\prime\prime}$ one can
show that after integration, the first term in expression
(\ref{neweq2}) will not contribute, while the second term
transpires to $c_0^2\kappa_0 f\left[q(t)\right]
f^\prime\left[q(t)\right]\epsilon(t)$.

Taking into consideration all the above assumptions and assuming
that the system variables evolve much more slowly in comparison to
the external noise $\epsilon(t)$, in the limit $\tau_c \rightarrow
0$, Eq.(\ref{eq12}) reduces to
\begin{eqnarray}\label{eq23}
\dot q &=& p, \nonumber \\
\dot p &=& - V^{\prime}(q) - \gamma
[f^{\prime}(q)]^2 p + f^{\prime}(q) \xi(t) + f^{\prime}(q) \pi(t) \nonumber \\
&& + \gamma \kappa_0 f(q) f^{\prime}(q) \epsilon(t) ,
\end{eqnarray}

\noindent where the dressed noise $\pi(t)$ obeys the correlation function
$\langle \pi(t) \pi(t^{\prime}) \rangle = 2 \gamma \kappa_0^2
\delta(t-t^{\prime})$.
Now for linear system-bath coupling, i.e., for $f(q)=q$, we obtain,
from Eq. (\ref{eq23}), the Langevin equation for the Brownian particle
\begin{eqnarray}\label{eq25}
\dot q &=& p \nonumber \\  \dot p &=& - V^{\prime}(q) - \gamma p +
\xi(t) + \pi(t) + q \eta(t) .
\end{eqnarray}

\noindent with $\eta(t) = \gamma \kappa_0 \epsilon(t)$. The noise
$\xi(t)$ is thermal noise and makes its presence felt due to the
coupling of the system with the reservoir. The other two noises,
$\pi(t)$ and $\eta(t)$, appear additively and multiplicatively,
respectively. As the origin of these two noise processes is the
same, that is the driving of the bath modes by $\epsilon(t)$, we
expect a cross correlation between them. Thus, it is easy to show
that
\begin{eqnarray}\label{eq27}
\langle \pi(t) \eta(t^{\prime}) \rangle =c_0 \kappa_0
\delta(t-t^{\prime}) = \langle \eta(t) \pi(t^{\prime}) \rangle .
\end{eqnarray}
\noindent Eq.(\ref{eq27}) reveals the fact that the two noise
processes $\pi(t)$ and $\eta(t)$ are mutually correlated with
$\kappa_0$ playing the role of degree of correlation. In writing
Eq.(\ref{eq27}) we redefine the noise $\pi(t)$ by absorbing the
negative sign.

From the very mode of the present development it is clear that we
use the paradigm of a hierarchy of interacting Hamiltonians which
underlies the reduced-dimensional nonequilibrium Langevin
equations so as to highlight which interaction is the driver for
space and time dependent correlations. Using an analytical
treatment of the system $+$ reservoir-bath $+$ driving-bath
hierarchy, we are able to demonstrate that the origin for the
nonstationary colored noise is the nonlinearity in the driving
term.

In this context we want to mention some allied formalisms made by
Hernandez and his group which bears a close kinship with our
present development. Very recently Popov and
Hernandez\cite{hernandez2} used a hierarchy of interacting
Hamiltonians to address the question of multiple temperature
baths, and the analytical approach used is similar to that taken
here. In this development they have addressed a generalized
construction for the effective temperature of a tagged particle
connected to an arbitrary number of time-dependent inhomogeneous
reservoirs. In a number of papers \cite{hernandez13} Hernandez and
co-workers explored the extent to which different molecular scale
perturbations can drive molecular systems to exhibit dynamic and
metastable properties accessible only within nonequilibrium
conditions. In these development they have also provided an
illustration of the nonstationary dissipation through the
justification for the use of the hierarchy of Hamiltonians
analytically. In this context Hernandez and co-workers
\cite{hernandez4,hernandez5} have also discussed treatments of an
external driving potential which lead to a rephrasing of the
frictional terms.

\section{Conclusion}

In conclusion,using the microscopic Hamiltonian picture we have
constructed a Langevin equation that apart from including the
thermal noise, originating from the heat bath, includes two
external mutually correlated noises, one appears additively and
the other multiplicatively. In almost all of the traditional
works, the external cross-correlated noises are treated
phenomenologically. Our approach is an attempt to understand the
underlying mechanism of additive and multiplicative cross
correlated noises and to realize their mutual interplay from a
microscopic view point.

\section{Acknowledgements}
We would like to thank the anonymous
reviewer for critical reading of our paper and various
critical suggestions. Financial support from CSIR, India [01(2257)/08/EMR-II] is
thankfully acknowledged.

\end{document}